\documentstyle[epsfig]{aipproc}
\begin{document}
\title{First-principles and semiempirical Hartree-Fock 
calculations for $F$ centers in KNbO$_3$ and Li
impurities in KTaO$_3$}
\author{R.~I.~Eglitis$^{a,b}$, E.~A.~Kotomin$^{b,c}$,    
A.~V.~Postnikov$^a$, N.~E.~Christensen$^c$, and 
G.~Borstel$^a$}
\address{
$^a$Universit\"at Osnabr\"uck -- Fachbereich Physik,   
D-49069 Osnabr\"uck, Germany \\ 
$^b$Institute of Solid State Physics, University of Latvia, 
8 Kengaraga, Riga LV-1063, Latvia \\
$^c$Institute of Physics and Astronomy, 
University of Aarhus  
Aarhus C, DK-8000, Denmark} 
\date{\centerline{(Received 4 February 1998)}}
\maketitle
\begin{abstract}
The linear muffin-tin-orbital method based on the density-functional
theory and the semi-empirical method of the
Intermediate Neglect of the Differential Overlap based on 
the Hartree--Fock formalism are used for the supercell study of
the $F$ centers  in cubic and orthorhombic
ferroelectric KNbO$_3$ crystals. Two electrons are found 
to be considerably delocalized even in the ground state of the defect.
The absorption energies were calculated by means of the INDO method using
the $\Delta$SCF scheme after a relaxation of atoms surrounding the $F$ center.

As an example of another type of point defect in perovskite,
an isolated Li impurity in KTaO$_3$ as well as interacting Li pairs
are considered in the supercell approach, using the supercells
of up to 270 atoms. The off-center Li displacement, reorientational energy
barriers and the lattice relaxation around impurities are calculated.
The results are compared with those obtained earlier within the shell
model, revealing the relaxation pattern somehow different from
the shell model estimations.
\end{abstract}

\section*{Introduction}
\label{sec:intro}
It is well understood now that point defects play an important role in
the electro-optic and non-linear optical applications of KNbO$_3$ and
related materials \cite{gun1}. In particular, the light
frequency doubling in KNbO$_3$ is seriously affected by unidentified defects
responsible for the induced IR absorption \cite{polzik}. The photorefractive
effect, important in particular for holographic storage, is also known
to depend on the presence of impurities and defects.

One of the most common defects in oxide crystals is the so-called $F$ center,
an O vacancy $V_{\mbox{\small O}}$ which traps two electrons \cite{craw}.
In electron-irradiated KNbO$_3$, a broad absorption band is observed 
around 2.7 eV at room temperature and tentatively ascribed to 
$F$ centers \cite{hod,F+}. The motivation behind the present study
of this defect by means of theory is twofold. 
First, due to the lowering of the local point symmetry at the O site
in ferroelectric phases of KNbO$_3$ (for instance, in the orthorhombic
phase which is the room-temperature one), the degeneracy of the 
$2p$-type excited state may be lifted, resulting in several splitted
absorption bands. Second, there were claims in the literature
in favor of the symmetry breaking between two Nb atoms neighboring
the O vacancy, resulting in an asymmetric electron density 
distribution \cite{pros}.
In order to clarify these questions and to check the assignment
of the 2.7 eV absorption band, we study in the present paper 
the $F$ center in KNbO$_3$ using the supercell model and two different 
theoretical techniques: the full-potential linear muffin-tin orbital
(FP-LMTO) and the semiempirical Intermediate Neglect 
of the Differential Overlap (INDO) methods. 
In the present contribution we essentially summarize
the results presented recently in more detail in Ref.~\cite{F-cent}.

The same techniques, also in the supercell approach, have been
applied to the study of another point defect, inducing lattice
relaxation in a incipient ferroelectric crystal: Li substituting
K in KTaO$_3$. As is experimentally known \cite{yac74},
such substitutional impurity gets spontaneously displaced along one 
of six possible [100]-type directions. The magnitude of this
displacement and (in some cases) the lattice relaxation related
to it have been estimated by empirical models \cite{Liempir,vdkk84}, 
the shell model \cite{stac90,ex:Li}, first-principles FP-LMTO 
calculations \cite{wil94} and recently by the INDO method \cite{KTL}.
As an extension of the latter study, we discuss in the present
contribution also the effects of interaction between Li impurities
by the same method. The INDO method has an advantage of being
relatively compact in what regards the necessary computer resources.
This allows in the study of impurity systems to treat 
relatively large supercells
which remain well beyond the range of {\it ab initio} total-energy
methods. At the same time, once the essential parameters of the
INDO method are known for the system in question, the description
of excitation energies and/or total-energy trends with the INDO
method is much better than within any model schemes.

The INDO method is described in detail in Ref.~\cite{indo1-3}.
We used its practical implementation in the computer code
CLUSTERD. The technical details related to the application
of method (e.g. parametrisation) for KNbO$_3$ can be found
in Ref.~\cite{indo}, for Li-doped KTaO$_3$ -- in Ref.~\cite{KTL}.
As a benchmark for the INDO parametrisation in the previous study
\cite{indo,KTL}, as well as for independent investigation of
the $F$-center in Ref.~\cite{F-cent} (the essential results of
which are discussed below), we used the FP-LMTO method
in the implementation by M.~Methfessel \cite{msm}. 

\section*{Technical details of calculation and results for F-centers}

In both LMTO and INDO calculations of the oxygen vacancy in KNbO$_3$, 
we used $2\!\times\!2\!\times\!2$ supercells, including 39 atoms, 
for the geometry of ideal cubic perovskite lattice.
A more detailed study of $F$-center system within the local
density approximation, using e.g. the atomic sphere approximation
along with the FP-LMTO, the electronic structure of the defect system
and the aspects of correcting the band gap are discussed
in Ref.~\cite{F-cent} and skipped here. 
Summarizing, the band gap estimated from the ground-state band
structure in the local density approximation (LDA), as is not justified
but commonly used, is known to be underestimated in dielectrics.
This does not present a problem for the total-energy studies
(even structure optimizations) in pure oxides, but in case
of the $F$-center the impurity band formed in the band gap
exhibits too strong dispersion (due to limited supercell size)
and overlaps with the states in the conduction band, resulting
in a metallic behavior of the impurity system. In the INDO
calculation which is essentially a Hartree-Fock scheme and hence
tends to overestimate the band gap, no such problem arises.
On the other hand, we would like to keep the LDA results as
a useful reference point for the electronic structure and
as a reliable benchmark of the total energy-based structure
optimization. It can be achieved either by an artificial operator
shifting the Nb$4d$ states in the conduction band upwards
(that is useful for the analysis of the electronic structure
but makes the total energy results unreliable), or by performing
the total energy calculation with only one {\bf k}-point in
the Brillouin zone, in order to suppress the dispersion
of the defect states. In the latter case, the calculation setup
becomes somehow resembling that of INDO, where also only
the $\Gamma$ point is traditionally used for the {\bf k}-space
sampling, in the spirit of the ``large unit cell'' (LUC) 
scheme \cite{evar}. Since the structure optimization normally
needs good convergency in the number of {\bf k}-points,
the total-energy result of such LDA calculation should be
considered as an approximate one, giving rather an error bar
when taken together with that of INDO. According to the LDA calculation,
the relaxation of two Nb atoms neighboring to the $F$ center
is outwards by $4.8\%$ of the lattice constant, resulting
in the energy lowering by 1.2 eV. From the INDO calculation,
both values are somehow larger but in qualitative agreement with the
LDA results: outward relaxation of two Nb neighbors by $6.5\%$
and the energy gain of $\sim$3.7 eV.

In addition to analyzing this most important aspect of relaxation,
we optimized in the INDO calculation the positions of more distant
neighbors (14 atoms in total) to the O vacancy as well.
The $0.9\%$ outward displacement of K atoms and
the $1.9\%$ inward displacement of O atoms gives the total
relaxation energy of $\sim$4.7 eV, mostly due to the contributions
from the Nb and O displacements.

The analysis of the effective charges of atoms surrounding the $F$ center
shows that of the two electrons associated with the removed O atom,
only $\approx -0.6 |e|$ is localized at the vacancy, and
about a similar extra charge is localized on the two nearest Nb atoms.
The $F$ center produces a local energy level $\approx 0.6$ eV
above the top of the valence band. Its molecular orbital contains primarily
contribution from the atomic orbitals of the two nearest Nb atoms.

\begin{table*}
\caption{Calculated absorption energy for the $F$ center ($E_{abs}$)
and the energy of the nearest-neighbors Nb relaxation ($E_{rel}$) 
in cubic and orthorhombic phases of KNbO$_3$.}
\label{tab:ene}
\begin{tabular}{l@{\hspace*{1.0cm}}ddd@{\hspace*{1.0cm}}d}
 Symmetry, phase       &   & $E_{abs}$ (eV) &    & $E_{rel}$ (eV) \\
\tableline
$C_{4v}$, cubic        & 2.73 & 2.97 &  --  & 3.7 \\
$C_s$, orthorhombic    & 2.56 & 3.03 & 3.10 & 3.6 \\
$C_{2v}$, orthorhombic & 2.72 & 3.04 & 3.11 & 3.6 \\
\end{tabular}
\end{table*}

The structure optimization based on the INDO calculation has also
been done for a low-symmetry geometry corresponding to a room-temperature
(ferroelectric) orthorhombic phase of KNbO$_3$.
This phase is stable in a broad temperature range (263 to 498 K)
and hence subject to most studies and practical applications.
The structure parameters in pure KNbO$_3$ have been optimized
with the INDO method in Ref.~\cite{indo} and are in quite good
agreement with the experimental measurements. In the orthorhombic phase,
there are two inequivalent positions of oxygen and hence the possibility
to form two different kinds of $F$-center, possibly with different
optical properties. The displacements of Nb atoms 
nearest to $V_{\mbox{\small O}}$ were calculated
for these both types of defects and found to be very close to those
found for the cubic phase. The relevant relaxation energies 
(3.6 eV) and also nearly the same as for the Nb relaxation 
found in the cubic phase.

Because of different local symmetry at the O site (either
$C_{2v}$ or $C_s$, in contrast to $C_{4h}$ in the cubic phase),
the energetics of the impurity levels changes.
In the cubic phase, the $V_{\mbox{\small O}}$ excited state splits 
into two levels, one of which remains two-fold degenerate. 
Our $\Delta$SCF calculations predict the two relevant
vacant bands, the absorption energies of which are given 
in Table \ref{tab:ene}.
In the orthorhombic phase, the degeneracy of the impurity level
is completely lifted (Table \ref{tab:ene}). The relaxation energies 
(associated with the Displacement of Nb neighbors only) are also
listed in Table \ref{tab:ene} for comparison.

\section*{Off-center displacement of L\lowercase{i}
in KT\lowercase{a}O$_3$ and related lattice relaxation}

\begin{figure}[t]
\centerline{\epsfig{file=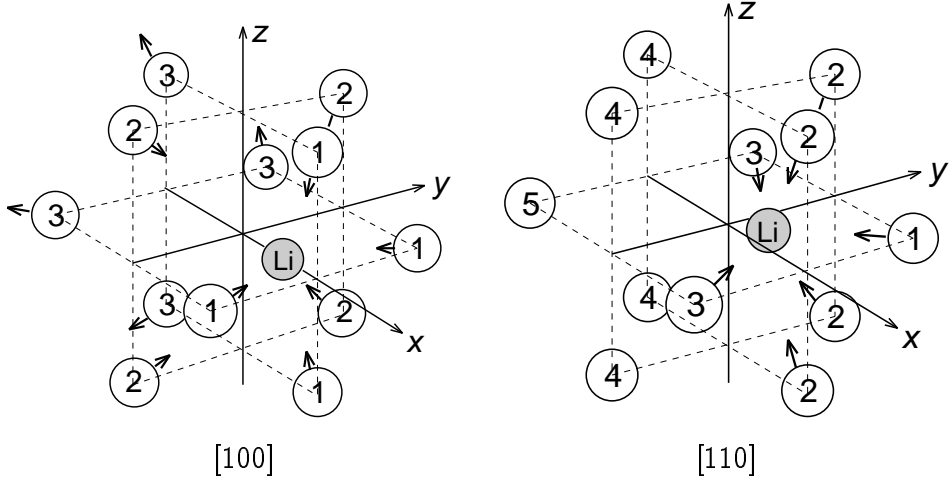,width=12.5cm}}
\vspace{10pt}
\caption{[100] and [110] Li off-center displacements
and the relaxation pattern of neighboring oxygen atoms.}
\label{fig:struc}
\end{figure}

The off-center displacement of substitutional Li in KTaO$_3$
is known to induce a considerable long-range polarization of crystal.
Therefore the supercell size needs to be larger than for the study
of $F$ centers. 
The size of the polarized region associated with the [100]-displaced
Li ion was estimated in the shell model calculation by Stachiotti
and Migoni\cite{stac90} to be about 5 lattice constants
along the direction of displacement, with $\sim$99\% of the
`effective dipole' polarization being confined to nearest
Ta--O chains, that go parallel to the displacement.
As is discussed below, the magnitudes of the atomic displacements
and polarization in our present calculation is considerably
smaller than those found in Ref.~\onlinecite{stac90}, and
the relaxed neighbors to the Li impurity are well within
the $3\!\times\!3\!\times\!3$ supercell. In order to be on safer side,
we performed as well the calculations for a supercell doubled
in the direction of Li displacement, i.e., $6\!\times\!3\!\times\!3$,
with a single [100]-displaced Li atom.
The equilibrium displacement in this case is 0.62~{\AA},
exactly as for the $3\!\times\!3\!\times\!3$ supercell,
with the energy lowering 57.2~meV. The difference
from the result for a $3\!\times\!3\!\times\!3$ supercell (62.0~meV)
roughly represents the uncertainty related to the supercell size
in our calculations.

\begin{figure}[tb]
\centerline{\epsfig{file=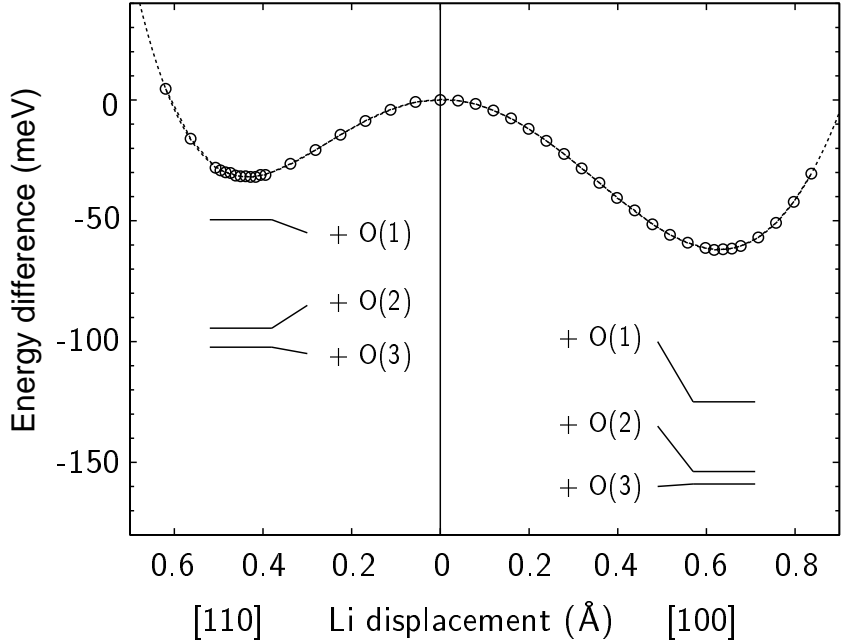,width=11.0cm}}
\vspace{10pt}
\caption{Total energy gain as function of [100] and [110] Li displacements
without lattice relaxation (dashed line with open circles),
and the total energy values after including the relaxation
of three groups of nearest oxygen atoms.}
\label{fig:relax}
\end{figure}

The parametrisation of the INDO method for the KTaO$_3$:Li system
has been done in Ref.~\onlinecite{KTL} based on the comparison
with the results of earlier FP-LMTO calculations \cite{wil94}
in what regards the magnitude of the Li off-center displacement
(0.61~{\AA}\cite{wil94}) as well as the energy gain due to
the Li displacement. The total energy as function of [100]
and [110] Li off-center displacements as calculated by INDO
is shown in Fig.~\ref{fig:relax}.
Our equilibrium [100] off-center displacement
of 1.44~{\AA} from the shell model by Stachiotti and Migoni
within the shell model\cite{stac90}.
On the other hand, our value is in good agreement with
a more recent, and apparently more elaborately parameterized, shell model
calculation by Exner {\it et al.} (0.64~{\AA}, Ref.~\onlinecite{ex:Li}).

\begin{table*}[b!]
\caption{Relaxed atomic positions for the Li displacement
as calculated by INDO.}
\label{tab:relax}
\begin{tabular}{ccccc}
Atom & \multicolumn{3}{c}{Lattice coordinates} & Displacement \\
\tableline
\multicolumn{5}{c}{Li [100] displacement} \\
           Li  & $\Delta_x$ & 0 & 0 & $\Delta_x=0.1550$ \\
 4$\times$O(1) & $\frac{1}{2}+\Delta_x$ & $\frac{1}{2}+\Delta_y$ & 0 &
 $\Delta_x=-0.0045$; $\Delta_y=-0.0105$ \\
 4$\times$O(2) & $\Delta_x$ & $\frac{1}{2}+\Delta_y$ &
			       $\frac{1}{2}+\Delta_y$ &
 $\Delta_x=0.0070$; $\Delta_y=-0.0026$  \\
 4$\times$O(3) & $-\frac{1}{2}+\Delta_x$ & $\frac{1}{2}+\Delta_y$ & 0 &
 $\Delta_x=-0.0020$; $\Delta_y= 0.0020$ 
\rule[-3mm]{0mm}{0mm} \\
\tableline
\multicolumn{5}{c}{Li [110] displacement} \\
           Li  & $\Delta_x$ & $\Delta_x$ & 0 & $\Delta_x=0.0760$ \\
 1$\times$O(1) & $\frac{1}{2}+\Delta_x$ & $\frac{1}{2}+\Delta_x$ & 0 &
 $\Delta_x=-0.0090$ \\
 4$\times$O(2) & $\frac{1}{2}+\Delta_x$ & $\Delta_y$ &
			       $\frac{1}{2}+\Delta_z$ &
 $\Delta_x=-0.0060$; $\Delta_y=0.0030$; $\Delta_z=-0.0080$  \\
 2$\times$O(3) & $\frac{1}{2}+\Delta_x$ & $-\frac{1}{2}+\Delta_y$ & 0 &
 $\Delta_x=-0.0020$; $\Delta_y=0.0060$ \\

 4$\times$O(4) & $-\frac{1}{2}+\Delta_x$ & $\Delta_y$ &
			       $-\frac{1}{2}+\Delta_z$ &
 $\Delta_x=0.0003, \Delta_y=0.0001, \Delta_z=0.0003$  \\

 1$\times$O(5) & $-\frac{1}{2}+\Delta_x$ & $-\frac{1}{2}+\Delta_x$ & 0 &
 $\Delta_x \sim 0$ 
\rule[-3mm]{0mm}{0mm} \\
\end{tabular}
\end{table*}

The displaced Li ion with its neighboring oxygen atoms
is shown schematically in Fig.~\ref{fig:struc} for the cases
of [100] and [110] off-center displacements. The [100] displacement
is known to occur in reality \cite{yac74} whereas the [110]
displacement represents a saddle point between two such adjacent
and equivalent displaced configurations.
Oxygen atoms are numbered according to their separation into
several inequivalent groups. 

The energy gain due to the Li off-center displacement
is not directly measurable in an experiment, but there are
estimations for the $90^0$-energy barrier via the saddle point
between [100] and [010]-displaced positions
to be 86~meV\cite{vdkk84}. 
Our estimate of the energy difference between [100] and [110] minima
is $\sim$30.2~meV, roughly two times larger than in the FP-LMTO
calculation\cite{wil94}, but much less than the experimental
estimate. The origin of this discrepancy, as has been mentioned
in Ref.~\onlinecite{wil94}, is most probably related to the lattice relaxation
around the displaced Li ion, that makes the net energy gain
from the displacement larger, and the $90^0$-activation energy
(involving now the displacement of many atoms) correspondingly higher.
Indeed, the second harmonic generation-based estimates of the activation
barrier\cite{voigt} reveal two types of processes, apparently
one involving the lattice relaxation (with the barrier height 86.2~meV)
and another one that is too fast for the lattice to follow,
with the barrier 14.7~meV.

In order to clarify this point, we performed a lattice relaxation
of several shells of neighbors to the displaced Li ion, for
the cases of [100] and [110] displacements. The relaxed
coordinates of atoms are given in Table \ref{tab:relax},
where the oxygen atoms are numbered
consistently with Fig.~\ref{fig:struc}.
The total energy values resulting
from the gradual inclusion of neighbor relaxation are shown
in Fig.~\ref{fig:relax}. We found the relaxation of twelve
nearest oxygen atoms essential, and the effect of relaxing nearest Ta
and more distant atoms to be negligible, in what regards the
effect on the total energy.
The energy gain in the fully relaxed [100]-displaced configuration,
with respect to a non-relaxed central Li position, is 158.9~meV;
the energy gain in the relaxed [110]-configuration is 102.3~meV.
Therefore, the enhancement of the excitation barrier due to
relaxation effects is by a factor of two, but still not
sufficient to reach experimentally expected $\sim$86 meV.
This discrepancy may be due to the fact that in reality
the $90^0$-reorientation process of the impurity does not
necessarily occur via the fully relaxed saddle-point configuration.
Depending on the actual degree of relaxation around the saddle-point
Li position, the barrier height is expected from Fig.~\ref{fig:relax}
to be between $\sim$57 meV (full relaxation at the saddle point)
to $\sim$127 meV (no relaxation).

\section*{Interacting L\lowercase{i} impurities 
in KT\lowercase{a}O$_3$}

The experimental investigations of the diluted K$_{1-x}$Li$_x$TaO$_3$ 
system are numerous and include e.g. the nuclear magnetic resonance
studies of relaxational dynamics associated with 
dipole reorientations \cite{vk12}, 
ultrasound attenuation measurements \cite{dou} and the measurements
of the low-frequency shear modulus\cite{hoc3}. 
Due to different technical limitations, none of these methods 
allows to attain the ground state of system in the concentration range 
$x\leq 7\%$. 

Up to now, there are reported only few theoretical studies
of interacting Li impurities in KTaO$_3$, using mainly 
analytical approaches\cite{vg90} or oversimplified shell
model calculations\cite{koh}, but there are no {\it ab initio}
studies reported to our knowledge at this field. In order to
get some theoretical predictions, check 
results of shell model calculations\cite{koh}, and answer 
the question about the  nature of the low-temperature
phase of K$_{1-x}$Li$_x$TaO$_3$ at small concentrations 
of Li spin glasses or ferroelectrics, INDO calculations
of Li-Li interaction in KTaO$_3$ may be of certain interest.

\begin{figure}[t]
\centerline{\epsfig{file=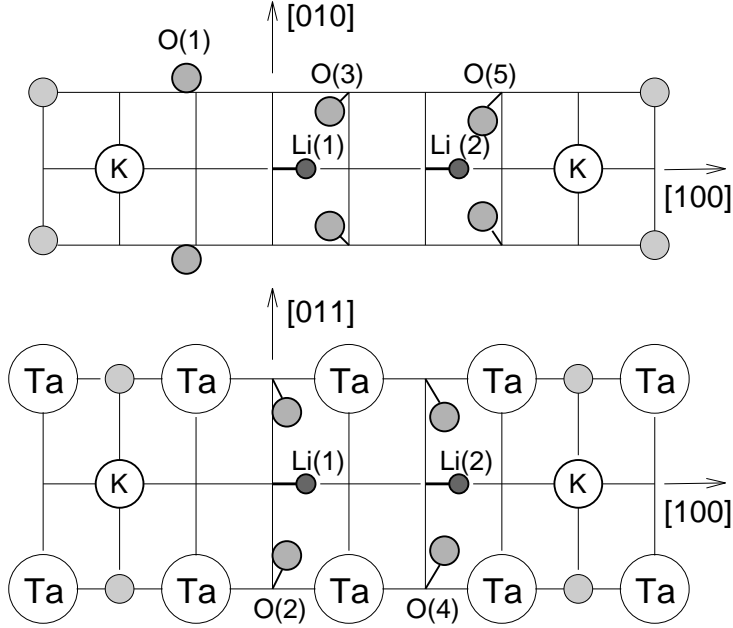,width=9.6cm}}
\vspace{10pt}
\caption{Off-center displacements of two nearest Li atoms
and the relaxation pattern of 20 neighboring oxygen atoms.}
\label{fig:relax2}
\end{figure}

In the preliminary calculations done by now, we concentrated on
two following subjects. First, we wanted to know how the interaction
between Li impurities which substitute two neighboring K sites
affects the energy characteristics and the lattice relaxation
associated with each impurity. For this purpose, we allowed
the simultaneous adjustment of the structure coordinates as listed
in Table \ref{tab:relax2} (affecting both impurities and 20
oxygen neighbors). The labelling of atoms in Tab.~\ref{tab:relax2}
and the qualitative scheme of the relaxation pattern is shown
in Fig.~\ref{fig:relax2}.
The relaxation of Ta and K atoms was found to be much smaller
than that of O neighbors. The energy gain that was 62 meV
due to the [100] displacement of a single Li impurity and
159 meV for the oxygen relaxation taken into account,
makes correspondingly 176 meV and 407 meV per two Li impurities.
It indicates that a substantial Li-Li interaction of the magnitude
$E(\mbox{2Li})-2\times E(\mbox{Li})$= 52 meV for bare Li
and 89 meV for Li with oxygen ``cloud'' indeed occurs and
is enhanced by lattice polarization.

\begin{table}[b]
\caption{
Lattice relaxation around two nearest Li impurities 
}
\label{tab:relax2}
\begin{tabular}{ccccc}
Atom & \multicolumn{3}{c}{Lattice coordinates} & Displacement \\
\hline
 Li(1)  & $\Delta_x$ & 0 & 0 & $\Delta_x=0.1550$ \rule[-0mm]{0mm}{8mm} \\
 Li(2) & 1+$\Delta_x$ & 0 & 0 & $\Delta_x=0.178$ \\
 O(1) &$-\frac{1}{2}+\Delta_x$ &$\pm$($\frac{1}{2}+\Delta_{yz}$) & 0 &
 $\Delta_x=-0.0022$; $\Delta_{yz}=0.0022$ \\
 O(2) & $\Delta_x$ &$\pm$($\frac{1}{2}+\Delta_{yz}$) &
 $\pm$($\frac{1}{2}+\Delta_{yz}$) &
 $\Delta_x=0.0075$; $\Delta_{yz}=-0.0028$  \\
O(3) & $\frac{1}{2}+\Delta_x$ &$\pm$($\frac{1}{2}+\Delta_{yz}$) & 0 &
 $\Delta_x=-0.0074$; $\Delta_{yz}=-0.0082$ \\
 O(4) & 1+$\Delta_x$ &$\pm$($\frac{1}{2}+\Delta_{yz}$) &
  $\pm$($\frac{1}{2}+\Delta_{yz}$) &
   $\Delta_x=0.0080$; $\Delta_{yz}=-0.0032$  \\
O(5) & $\frac{3}{2}+\Delta_x$ &$\pm$($\frac{1}{2}+\Delta_{yz}$) & 0 &
 $\Delta_x=-0.0057$; $\Delta_{yz}=-0.0125$ 
\rule[-3mm]{0mm}{0mm} \\
\end{tabular}
\end{table}

\begin{figure}
\epsfxsize=8.5cm
\centerline{\epsfig{file=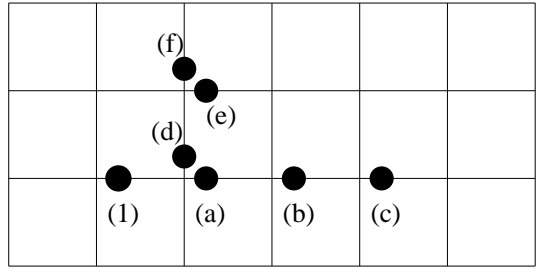,width=7.0cm}}
\vspace{10pt}
\caption{Distribution of interacting Li impurities 
in 3$\times$3$\times$6 extended KTaO$_3$ supercell. 
There are considered interaction between Li(1)
and another Li placed consequently in a,b,c,d,e,f 
positions.}
\label{fig:LiLi}
\end{figure}

Second important characteristic is the strength and spatial distribution
of the dipole field created by each Li impurity. In order to study it,
we used the ``probe'' Li impurity at different lattice positions
and displaced in different directions with respect to the ``central''
one. The ``central'' impurity was displaced along [100] by 0.62 {\AA},
i.e. the equilibrium displacement for the single off-center Li ion.
The ``probe'' impurity was allowed to relax along the given direction,
and the interaction energy was extracted as a measure of the dipole
field in crystal.
The tested positions of the ``probe'' impurities in crystal
are indicated in Fig.~\ref{fig:LiLi}, and the resulting interaction
energies -- in Table \ref{tab:ek3}. One can clearly see the
anisotropy of the interaction field. The numerical values may be
somehow affected by the choice of even larger supercell in
subsequent calculations. It is noteworthy, however, that the
interaction strength decreases with the distance faster than
it was found in the shell model calculations \cite{koh}.

\begin{table*}
\caption{Effective interaction energy $E_{int.}$ 
and equilibrium displacement $\Delta$ of the second Li atom (a--f)
for different mutual configurations displacements.
The displacement of Li(1) is fixed at 0.62 {\AA} along [100].}
 \begin{tabular}{cdd@{\hspace*{1.0cm}}cdd}
 Li pair & $E_{int}$ (meV) & $\Delta$ (\AA) &
 Li pair & $E_{int}$ (meV) & $\Delta$ (\AA) \\
\tableline
 1--a & 61.6  & 0.72  & 1--d & 28.52 & 0.677 \\
 1--b & 10.29 & 0.633 & 1--e &  5.93 & 0.633 \\
 1--c & 7.56  & 0.629 & 1--f & 17.91 & 0.657  \\
\end{tabular}
\label{tab:ek3}
\end{table*}

\section*{Summary}

In the present study of two different point defects in perovskites
by means of a semiempirical INDO method, based on the comparison
with {\it ab initio} calculation results, we established the following. 
The ground state of the $F$ center in KNbO$_3$ is associated with a strong 
symmetrical relaxation of two nearest Nb atoms outwards relative 
to the O vacancy.
We presented a strong argument that the 2.7 eV absorption band observed in
electron-irradiated crystals could be due to the $F$-type centers, and
predicted the existence of two additional absorption bands 
(at 3.04 eV and 3.10 eV) for the same defect in the orthorhombic phase 
of KNbO$_3$.
In the analysis of single and interacting off-center Li impurities
in KTaO$_3$ we estimated the lattice relaxation around impurities
and the characteristic interaction energies, depending on the distance
between the defects and their mutual orientation. The interaction
energies are lower and less long-ranged than it was estimated from
earlier shell-model calculation. \\

{\bf Acknowledgments}\\*[0.2cm]  
Financial support of the Deutsche Forschungsgemeinschaft
(SFB 225, Graduate College) is greatly acknowledged.


\begin{references}
\bibitem{gun1}    P.~G{\"u}nter and J.-P.~Huignard (eds.)
                  {\it Photorefractive Materials and Their Application} 
		  (Topics in Applied Physics, {\bf 61, 62}), 
		  Berlin, Heidelberg: Springer-Verlag, 1988.
\bibitem{polzik}  L.~Shiv, J.~L.~S{\o}rensen, E.~S.~Polzik, and G.~Mizell,
                  {\it Optics Letters} {\bf 20}, 2271 (1995).
\bibitem{craw}    J.~H.~Crawford, Jr., 
		  {\it Nucl. Inst. Meth. B} {\bf 1} 159 (1984);
                  J.~-M.~Spaeth, J.~R.~Niklas, and R.~H.~Bartram, 
		  {\it Structural Analysis of Point Defects in Solids}
                  (Springer series in Solid State Sciences, vol. 43),
		  Berlin, Heidelberg: Springer-Verlag, 1993.
\bibitem{hod}     E.~R.~Hodgson, C.~Zaldo, and F.~Agullo-L{\'o}pez, 
		  {\it Solid State Commun.} {\bf 75}, 351 (1990).
\bibitem{F+}      E.~A.~Kotomin, R.~I.~Eglitis, and A.~I.~Popov, 
		  {\it J.~Phys.: Condens.~Matter} {\bf 9}, L315 (1997).
\bibitem{pros}    S.~A.~Prosandeyev, A.~V.~Fisenko, A.~I.~Riabchinski, 
		  A.~I.~Osipenko, I.~P.~Raevski, and N.~Safontseva, 
		  {\it J.~Phys.: Condens.~Matter},{\bf 8}, 6705 (1996).
\bibitem{F-cent}  R.~I.~Eglitis, N.~E.~Christensen, E.~A.~Kotomin,
		  A.~V.~Postnikov, and G.~Borstel,
		  {\it Phys.~Rev.~B} {\bf 56}, 8599 (1997).
\bibitem{yac74}   Y.~Yacoby and S.~Just,
		  {\it Solid State Commun.} {\bf 15}, 715 (1974).
\bibitem{Liempir} F.~Borsa, U.~H\"ochli, J.~J.~van der Klink, and D.~Rytz,
		  {\it Phys.~Rev.~Lett.} {\bf 45}, 1884 (1980);
		  U.~T.~H\"ochli, K.~Knorr, and A.~Loidl,
		  {\it Adv.~Phys.} {\bf 39}, 405 (1990).
\bibitem{vdkk84}  J.~J.~van der Klink and S.~N.~Khanna,
		  {\it Phys.~Rev.~B} {\bf 29}, 2415 (1984).
\bibitem{stac90}  M.~G.~Stachiotti and R.~L.~Migoni,
		  {\it J.~Phys.: Condens.~Matter.} {\bf 2}, 4341 (1990).
\bibitem{ex:Li}   M.~Exner, C.~R.~A.~Catlow, H.~Donnerberg, and
                  O.~F.~Schirmer,
                  {\it J.~Phys.: Condens.~Matter} {\bf 6}, 3379 (1994).
\bibitem{wil94}   A.~V.~Postnikov, T.~Neumann, and G.~Borstel,
         	  {\it Ferroelectrics}, {\bf 164}, 101 (1995).
\bibitem{KTL}     R.~E.~Eglitis, A.~V.~Postnikov, and G.~Borstel,
		  {\it Phys.~Rev.~B} {\bf 55}, 12976 (1997).
\bibitem{indo1-3} A.~Shluger,
		  {\it Theoret.~Chim.~Acta} (Berl.) {\bf 66}, 355 (1985);
		  E.~Stefanovich, E.~Shidlovskaya, A.~Shluger,
		  and M.~Zakharov,
		  {\it Phys.~Status~Solidi B} {\bf 160}, 529 (1990);
                  A.~Shluger and E.~Stefanovich,
		  {\it Phys.~Rev.~B} {\bf 42}, 9664 (1990).
\bibitem{indo}    R.~I.~Eglitis, A.~V.~Postnikov, and G.~Borstel,
		  {\it Phys.~Rev.~B} {\bf 54}, 2421 (1996).
\bibitem{msm}     M.~Methfessel,
		  {\it Phys.~Rev.~B} {\bf 38}, 1537 (1988);
                  M.~Methfessel, C.~O.~Rodriguez, and O.~K.~Andersen,
		  {\it ibid.} {\bf 40}, 2009 (1989).
\bibitem{evar}    R.~A.~Evarestov and L.~A.~Lovchikov, 
                  {\it Phys.~Status Solidi B} {\bf 93}, 469 (1977).
\bibitem{voigt}   P.~Voigt and S.~Kapphan,
		  {\it J.~Phys.~Chem.~Solids} {\bf 55}, 853 (1994).
\bibitem{vk12}    J.~J.~van der Klink and F.Borsa,
		  {\it Phys.~Rev.~B} {\bf 30}, 52 (1992);
                  S.~Rod, F.~Borsa, and J.~J.~van der Klink,
		  {\it ibid.} {\bf 38}, 2267 (1988).
\bibitem{dou}     P.~Doussineau, C.~Frenois, A.~Lavelut, and S.~Ziolkievicz,
		  {\it J.~Phys.: Condens.~Matter} {\bf 3}, 8369 (1991).
\bibitem{hoc3}    U.~H\"ochli, J.~Hessinger, and K.~Knorr,
		  {\it J.~Phys.: Condens.~Matter} {\bf 3}, 8377 (1991).
\bibitem{vg90}    B.~E.~Vugmeister and M.~D.~Glinchuk,
		  {\it Rev.~Mod.~Phys.} {\bf 82}, 993 (1990).
\bibitem{koh}     M.~G.~Stachiotti, R.~L.~Migoni, H.~M.~Christen, 
		  J.~Kohanoff, and U.~H\"ochli, 
		  {\it J. Phys.: Condens. Matter} {\bf 6}, 4297 (1994).
\end{references}
\end{document}